\documentclass[twocolumn,superscriptaddress,pra]{revtex4-1}
\usepackage{graphicx}
\usepackage{amsmath}
\usepackage{amsthm}
\usepackage{amssymb}
\usepackage{latexsym}
\usepackage{array}
\usepackage{float}
\usepackage{amsfonts}
\usepackage{mathrsfs}
\usepackage{verbatim}

\usepackage{color}

\begin{document}

\title{Experimental Demonstration on Quantum Sensitivity to Available Information in Decision Making}

\author{Joong-Sung~Lee}
\affiliation{Department of Physics, Hanyang University, Seoul 04763, Korea}

\author{Jeongho~Bang}\email{jbang@kias.re.kr}
\affiliation{School of Computational Sciences, Korea Institute for Advanced Study, Seoul 02455, Korea}
\affiliation{Institute of Theoretical Physics and Astrophysics, University of Gda\'{n}sk, 80-952 Gda\'{n}sk, Poland}

\author{Jinhyoung~Lee}\email{hyoung@hanyang.ac.kr}
\affiliation{Department of Physics, Hanyang University, Seoul 04763, Korea}

\author{Kwang-Geol~Lee}\email{kglee@hanyang.ac.kr}
\affiliation{Department of Physics, Hanyang University, Seoul 04763, Korea}
\received{\today}

\begin{abstract}
We present an experimental illustration on the quantum sensitivity of decision making machinery. In the decision making process, we consider the role of available information, say hint, whether it influences the optimal choices. To the end, we consider a machinery method of decision making in a probabilistic way. Our main result shows that in decision making process our quantum machine is more highly sensitive than its classical counterpart to the hints we categorize into ``good'' and ``poor.'' This quantum feature originates from the quantum superposition involved in the decision making process. We also show that the quantum sensitivity persists before the quantum superposition is completely destroyed.
\end{abstract}

\maketitle

\newcommand{\bra}[1]{\left<#1\right|}
\newcommand{\ket}[1]{\left|#1\right>}
\newcommand{\abs}[1]{\left|#1\right|}
\newcommand{\expt}[1]{\left<#1\right>}
\newcommand{\braket}[2]{\left<{#1}|{#2}\right>}
\newcommand{\commt}[2]{\left[{#1},{#2}\right]}

\newcommand{\tr}[1]{\mbox{Tr}{#1}}

We live in a chain of decisions everyday. We make a decision whether to take an umbrella as assessing the chance of raining.  Decisions are made by accounting for available information, e.g., the dark clouds through a window and/or the 30\% chance of raining that the weather forecast announces.  Yet we often make wrong decisions due to inadequate or noisy information. The relations of decisions with given information were studied in the theory of decision making (DM)\cite{Zsambok14}. However, it is not easy that DM processes are consistently analyzed\cite{Tversky74,Tversky92}. This is mainly because each decision maker has the different degree of ``sensitivity'' to a given available information; ones are more biased with the given information than others\cite{Blackhart05,Zsambok14}. This is an intrinsic trait of decision makers\cite{Resulaj09}. In this work we focus on the sensitivity to the available information which we categorize as ``good'' and ``poor'' hints, qualitatively.

Our DM study is presented in a framework of game theory\cite{Julio10}. Game theory deals with the strategies by which players (decision makers in this paper) maximize their own rewards. Nowadays quantum science has extended game theory to the quantum domain, revealing distinctive quantum features and opening a new avenue of applications\cite{Meyer99,Eisert99,Lee03}. As in quantum game theory, we are to investigate a quantum trait in decision makers, which originates from quantum properties\cite{Deutsch99,Pothos09}, i.e. the quantum sensitivity to the available information during a DM process. This is intimate to an issue of quantum game theory, whether any quantum effects are revealed when no quantum strategies are involved. This has been regarded to be negative\cite{SJvanEnk02,Aharon08}. To this end, we consider machines which play (or simulate) rational decision makers, equipped with a simple and reasonable DM algorithm. We then compare the two types of decision making machines, classical and quantum. Here, game elements including strategies are assumed to be classical, except the decision processes, in which the quantum machine is allowed to exploit a quantum algorithm\cite{Bang16}. Our main result shows that the quantum decision maker is more highly sensitive than its classical counterpart to given available information, categorized to good and poor hints. This is attributed to the quantum coherence involved in the quantum DM process. We also show that the quantum sensitivity persists before the quantum coherence is completely destroyed. These results will be applicable to reinforcement learning and preference updating\cite{Mihatsch02,Lee08,Molleman14,Ghahramani15}; they expect a risk-averse machine to learn more slowly.

\begin{figure}
\centering
\includegraphics[width=0.46\textwidth]{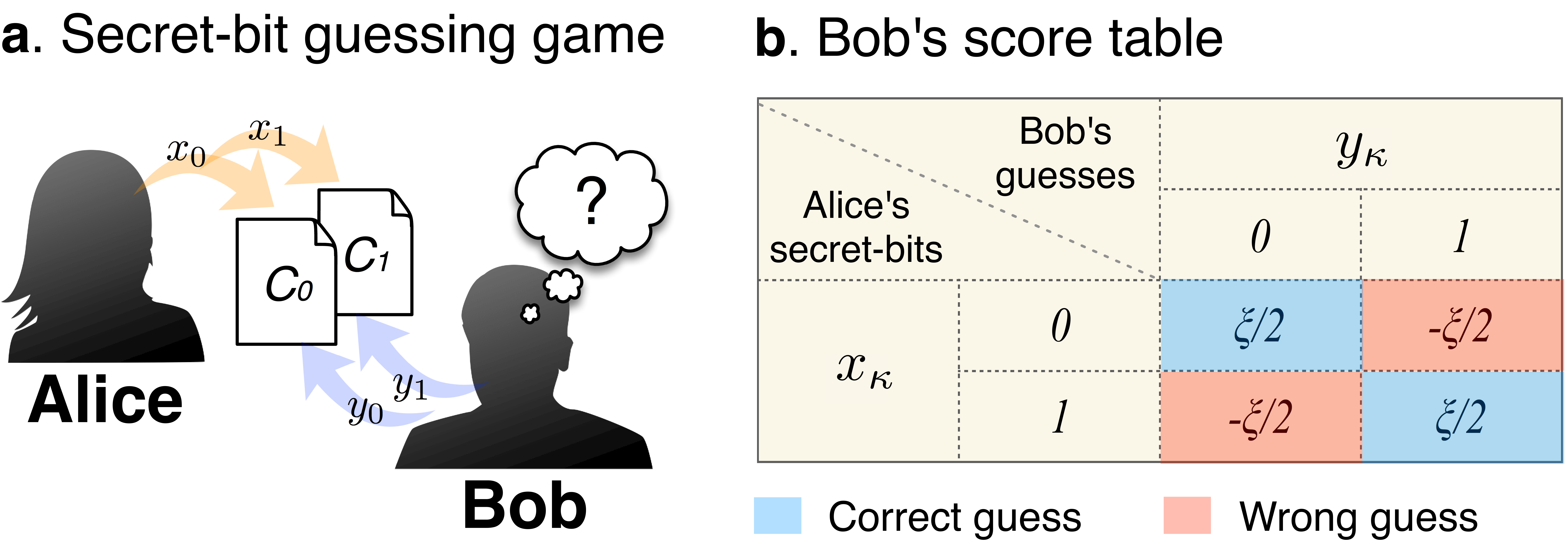}
\caption{\label{fig:game} {\bf Schematic picture of a secret-bit guessing game.} ({\bf a}) One player Bob guesses the numbers chosen by the other player, say Alice. Alice selects two numbers $x_\kappa \in \{0,1\}$ and writes on two cards $C_\kappa$. These numbers are unknown for Bob. Bob is to guess Alice's secret numbers $x_\kappa$. In doing so, Bob can exploit some available information, which we call ``hints.'' ({\bf b}) Table presents the scores which Bob will get in the game. Bob receives a score, positive of $\xi/2$ on a correct guess and negative of $-\xi/2$ on a wrong guess.}
\end{figure}

{\em Secret-bit guessing game.}---We suggest a simple game, called the ``secret-bit guessing game'' (see Fig.~\ref{fig:game}{\bf a})\cite{Lungo05}. In this game, one player (say Alice) has a couple of cards $C_\kappa$ ($\kappa=0,1$), on each of which her secret-bit number $x_\kappa$ is written. The other player (say Bob) should make a guess $y_\kappa$ (or ``strategy'' in the language of game theory) at her secret-bit $x_\kappa$. By a successful guess (i.e., $y_\kappa = x_\kappa$), Bob receives a positive score of $\xi/2$; however, by a wrong guess (i.e., $y_\kappa \neq x_\kappa$), Bob receives a penalty, i.e., a negative score of $-\xi/2$ (see Fig.~\ref{fig:game}{\bf b}). After the two guesses, Bob will get a score among $\{ -\xi, 0, \xi \}$. Then, Bob wins (loses) with a score of $\xi$ ($-\xi$). The game ends in a draw if Bob has a score of zero. Here, we raise a question whether some (additional) hints can help Bob to increase his winning probability or score. In particular, we explore how Bob's winning probability depends on a DM algorithm, considering the two types of DM processes which work classically and quantum-mechanically, respectively. Our results suggest that some quantum features play roles in the DM process with no use of quantum strategies.

\begin{figure}
\centering
\includegraphics[width=0.46\textwidth]{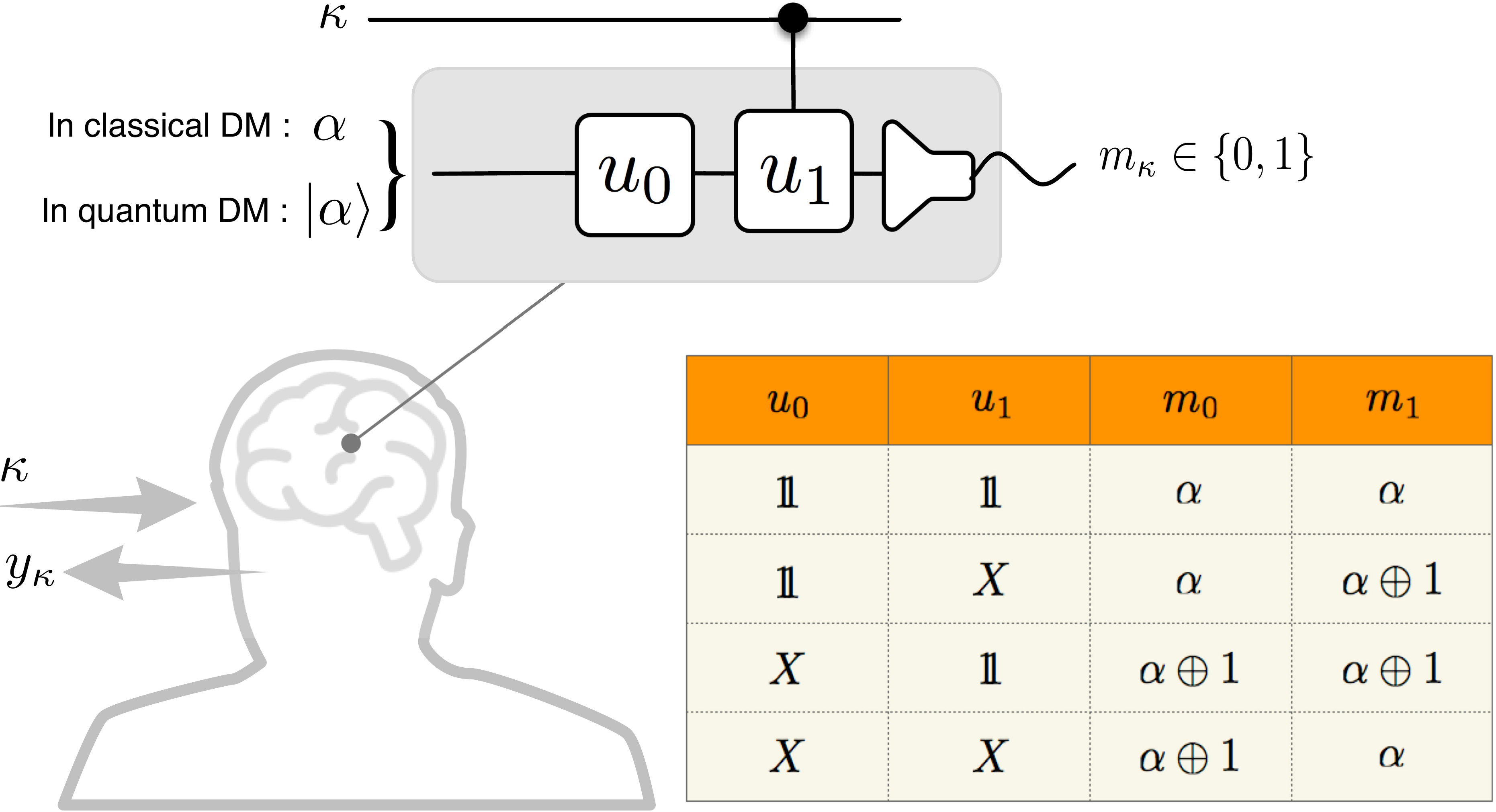}
\caption{\label{fig:dm_algorithm} {\bf Bob's decision making (DM) algorithm.} A machinery with an algorithm is assumed to simulate Bob's decision-making process. We consider and compare the machines of two types, classical and quantum. Equipped with a DM algorithm, machine ``Bob'' is supposed to guess Alice's secret-bit number $x_\kappa$ on card $C_\kappa$ for each input $\kappa \in \{0,1\}$. The algorithm implements all possible guesses of Bob with two operations $u_{0}$ and $u_{1}$ in the ancillary channel. The operation $u_1$ is conditional on input $\kappa$: i.e., $u_1$ is applied only if $\kappa=1$.  In case of the quantum machine, the operations $u_{0}$ and $u_{1}$ are unitary, applied to an initial fiducial state $\ket{\alpha}$, where each of them is composed of quantum superpositions with the identity (doing nothing) and the logical-not (flipping). The output state in the ancillary channel is measured with outcome $m_\kappa \in \{0,1\}$. In case of the classical counterpart, on the other hand, the operations are stochastic and work probabilistically the identity or logical-not, to the initial fiducial bit value $\alpha$, with outcome $m_\kappa$. Then, Bob's guesses at Alice's secret numbers $x_\kappa$ are given by $y_\kappa=m_\kappa \oplus \alpha$. Table lists the outcomes $m_\kappa$ generated by the possible set of operations $u_{0}$ and $u_{1}$ in the deterministic cases.}
\end{figure}

{\em Classical \& quantum decision making.}---To proceed, we adopt a DM algorithm, which is assumed to work in Bob's brain. The DM algorithm is modeled as a machinery process (see Fig.~\ref{fig:dm_algorithm}), which runs with two channels: an input channel of a single bit for Alice's card number $\kappa \in \{0,1\}$, and the other is an ancillary channel for processing the input with an output which is used for Bob's guess. The ancillary channel consists of two {\em probabilistic} operations $u_j$ ($j=0,1$), each supposed to be either the identity $\openone$ (doing nothing) or the logical-not $X$ (flipping the signal). Here, applying $u_1$ is conditioned on the input $\kappa$: i.e., $u_1$ is applied only if $\kappa=1$. The algorithm commences with receiving an input $\kappa$ from Alice. The two probabilistic operations $u_j$ in the ancillary channel are carried out with respect to the probabilities $P(u_j \to \openone)$ and $P(u_j \to X) = 1-P(u_j \to \openone)$. Here, $P(u_j \to \openone)$ and $P(u_j \to X)$ are the probabilities that $u_j$ is to be $\openone$ and $X$, respectively. The ancillary input is prepared to a fiducial bit $\alpha$ in the classical case or state $\ket{\alpha}$ in the quantum case. It is flipped or unchanged as successively passing through $u_0$ and $u_1$. The output is measured with an outcome $m_k \in \{0,1\}$. Then, Bob's guess $y_\kappa$ at Alice's secret numbers $x_\kappa$ is made such that $y_\kappa =  m_\kappa \oplus \alpha$ for each input $\kappa$. Note that this DM algorithm is universal in the sense that it realizes all possible guesses $y_\kappa$ of Bob (for more details, see Table in Fig.~\ref{fig:dm_algorithm} and/or Sec.~S1-A of the Supplementary Material).

Here the probabilities $P(u_j \to \openone)$ and $P(u_j \to X)$ ($j=0,1$) refer to the DM preferences\cite{Julio10}. For example, if $P(u_j \to \openone)$ is larger than $\frac{1}{2}$, Bob (or his brain) prefers setting $u_j \to \openone$ to $u_j \to X$. We can represent these probabilities as (for $j=0,1$)
\begin{eqnarray}
P(u_j \to \openone) = \frac{1}{2} + h_j~\text{and}~P(u_j \to X) = \frac{1}{2} - h_j,
\label{eq:preferences}
\end{eqnarray}
where hint $h_j \in [-\frac{1}{2}, \frac{1}{2}]$. Note that the hints are not always informative\cite{Lehner13}; for instance, a decision maker may acquire some hint fabricated with malicious, which we say poor. We thus need to characterize the quality of given hints, which we represent by a hint vector $\mathbf{h}=(h_0,h_1)^T$. We categorize hint vectors into ``good'' and ``poor.'' A hint vector $\mathbf{h}$ is categorized to good if, by using it, Bob can improve his winning probability. Otherwise, it is to poor. 

We consider and compare the machinery DM processes of two types, classical and quantum. The classical DM (cDM) is defined using the classical elements for the ancillary channel: the input $\alpha$ is a classical bit number and $u_j$ ($j=0,1$) is applied in a classical probabilistic way, namely, either to be $\openone$ or to be $X$ based on Eq.~(\ref{eq:preferences}). In this case, the probabilistic application of $u_j$ is represented by a stochastic evolution matrix,
\begin{eqnarray}
\begin{pmatrix}
P(u_j \to \openone) & P(u_j \to X) \\
P(u_j \to X) & P(u_j \to \openone)
\label{eq:sto_evol_mat}
\end{pmatrix}.
\end{eqnarray} 
On the other hand, the quantum DM (qDM) runs with the quantum state $\ket{\alpha}$ and the application of $u_j$ is represented by a unitary matrix,
\begin{eqnarray}
\begin{pmatrix}
\sqrt{P(u_j \to \openone)} & e^{i \phi_j}\sqrt{P(u_j \to X)} \\
e^{-i \phi_j}\sqrt{P(u_j \to X)} & -\sqrt{P(u_j \to \openone)}
\label{eq:unitary_mat}
\end{pmatrix}.
\end{eqnarray}
Here we note that the additional degree of freedom, i.e., the quantum phase $\phi_j$, is introduced in the unitary operation. The qDM utilizes these phases with the directional condition $\mathbf{h}=(h_0, h_1)^T$ in addition to the individual components of $\mathbf{h}$, according to the following rules:
\begin{eqnarray}
\left\{
\begin{array}{ll}
\Delta = 0 & \text{if}~h_0 h_1 > 0, \\
\Delta = \pi & \text{if}~h_0 h_1 < 0, \\
\Delta = \frac{\pi}{2} & \text{if}~h_0 h_1 = 0, 
\end{array}
\right.
\label{eq:qh_rule}
\end{eqnarray}
where $\Delta = \left| \phi_1 - \phi_0 \right|$ is defined as the absolute difference of the quantum phases $\phi_j$. These rules were built based on the postulate of ``rational'' game player (Bob, here) who can find the best algorithm by utilizing all available resources---which is often referred to as the theory of rationality\cite{Julio10}. Actually, the rules in Eq.~(\ref{eq:qh_rule}) optimizes Bob's DM algorithm and thus maximizes his winning probability (see Sec.~S1-B of the Supplementary Material). It is worth noting that we run the DM process quantum-mechanically, even though we keep the game strategies classical, such as Alice's secret numbers and Bob's guesses.

{\em Quantum sensitivity to additional hints.}---In such settings, we investigate quantum sensitivity to the given hints. First, we indicate that qDM allows Bob to enjoy much higher winnings with good hints. More specifically, by analyzing Bob's average score $\Xi$ (often-called the average payoff function --- a term from game theory)\cite{Julio10}, we arrive at
\begin{eqnarray}
\Xi_Q = \Xi_C + \Gamma, 
\end{eqnarray}
where the indices $C$ and $Q$ denote classical and quantum, respectively. Bob's quantum score differentiates from the classical by the amount of $\Gamma$. We set $\alpha=0$ and $\xi=1$ for a sake of simplicity. As in Eq.~(S15), the Supplementary Materials, the differential
\begin{eqnarray}
\Gamma = 2\sqrt{\left(\frac{1}{4}-h_0^2\right)\left(\frac{1}{4}-h_1^2\right)},
\end{eqnarray}
and clearly this leads to an advantage for qDM since $\Gamma \ge 0$. If the hints are poor, on the other hand, qDM makes it more difficult to make the correct guesses. In the worst case [see Eq.~(S16) in the Supplementary Materials],
\begin{eqnarray}
\Xi_Q = \Xi_C - \Gamma.
\end{eqnarray}
This implies that the differential $\Gamma$ becomes disadvantageous with the minus sign. Here, the most surprising fact is that, in qDM, Bob's score exhibits an abrupt transition near the boundary between good and poor hints. For example, when the amounts of hints are small but non-zero, approximately Bob's scores $\Xi_Q \simeq +\Gamma$ and $\Xi_Q \simeq -\Gamma$ for the good and poor hints, respectively, if the hints are symmetric, i.e., $\abs{h_0} = \abs{h_1} = \abs{h}$, where the symmetric hints were taken into account as hints are usually dependent and correlated. As the symmetric hint comes to zero, more explicitly, Bob's quantum score
\begin{eqnarray}
\Xi_Q \to
\left\{
\begin{array}{l}
+\Gamma \simeq +0.5~\text{as}~\abs{h^{(G)}} \to 0, \\
-\Gamma \simeq -0.5~\text{as}~\abs{h^{(P)}} \to 0,
\end{array}
\right.
\label{eq:qpt}
\end{eqnarray}
where we used $\Xi_C \to 0$ as $\abs{h^{(G,P)}} \to 0$. Here, $h^{(G)}$ and $h^{(P)}$ respectively stand for the good and poor symmetric hints. This abrupt score-transition (which resembles quantum phase transition)\cite{Park16} is a representative of the quanum sensitivity. Without any hints, i.e., $\abs{\mathbf{h}}=0$, however, there is no gain or loss from the quantum assumption (for detailed calculations and theoretical analyses, see Sec. S1-B of the Supplementary Material). 

\begin{figure}
\centering
\includegraphics[width=0.46\textwidth]{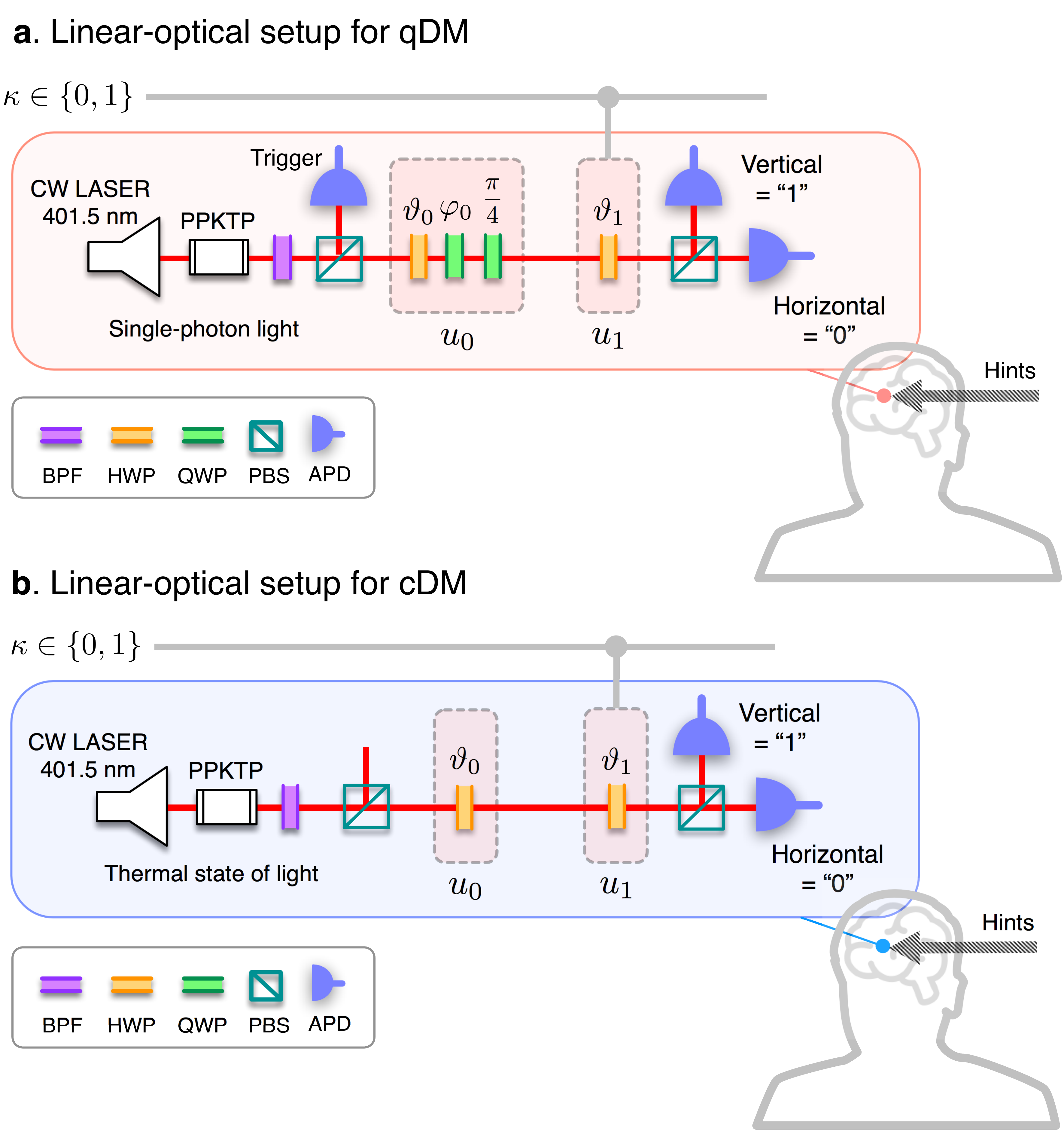}
\caption{\label{fig:exp} {\bf Linear-optical setups for simulating two types of DM, classical (cDM) and quantum (qDM).} ({\bf a}) In qDM, heralded single photons are prepared as the input light source by applying the post-selection to orthogonally polarized photon pairs generated by a type-II SPDC process (see Methods for more details). The single-photon polarizations, i.e., horizontal ($H$) and vertical ($V$), are employed as a quantum bit (qubit), an information carrier in the ancillary channel. The first operation $u_0$ is composed of HWP($\vartheta_0$)-QWP($\varphi_0$)-QWP($\pi/4$) with the controlling angles $\vartheta_0$ and $\varphi_0$, where HWP and QWP are half and quarter wave plates. The second operation $u_1$ is realized by only HWP($\vartheta_1$) with the controlling angle $\vartheta_1$. In this setting, $u_0$ and $u_1$ are so adjusted according to the rules in Eq.~(\ref{eq:control_rule_qDM}) together with Eq.~(\ref{eq:preferences}). The quantum interference between the two unitary operations of the single-photon polarization is thus exploited in the qDM. ({\bf b}) In cDM, the thermal state of light is employed as the ancillary input, which does not possess the quantum coherence. To do so, we do not apply the post-selection contrary to the qDM. The operations $u_{0,1}$ are implemented by only HWPs with either $\vartheta_j=0$ (for $u_j \to$ identity $\openone$) or $\vartheta_j=\frac{\pi}{4}$ (for $u_j \to$ logical-not $X$), randomly chosen in the probabilities by Eq.~(\ref{eq:preferences}).}
\end{figure}

\begin{figure*}
\centering
\includegraphics[width=1.00\textwidth]{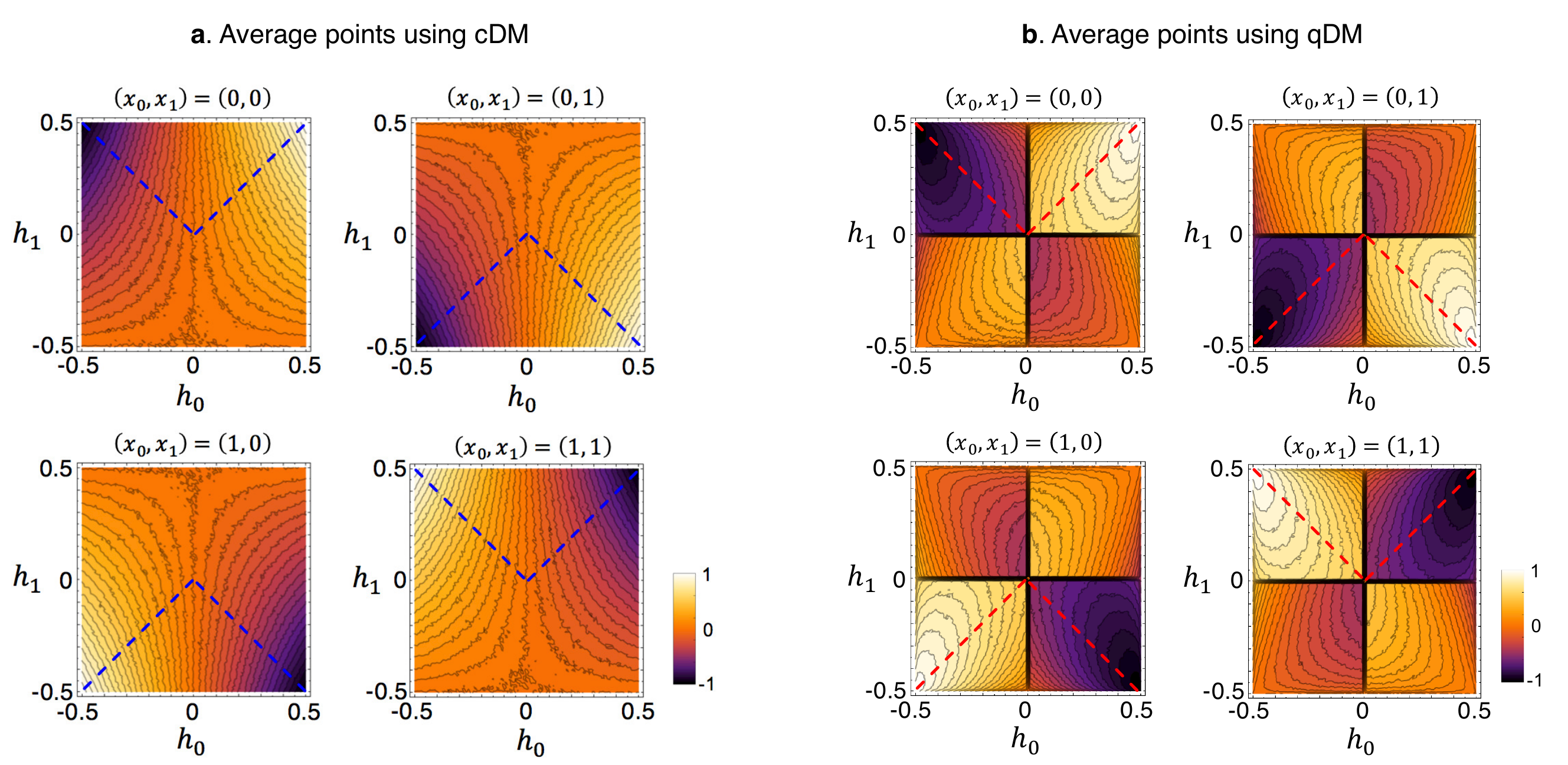} 
\caption{\label{grp:exp_results1} {\bf Average scores for Bob in the cDM and the qDM experiments.} Bob's average scores for all pairs of Alice's secret bits ($x_0$, $x_1$) are presented in the density plots for ({\bf a}) cDM and ({\bf b}) qDM experiments as described in Fig.~\ref{fig:exp}. The score values are obtained by repeating $10^4$ games for each hint vector $\mathbf{h} = (h_0, h_1)^T$, on the square mesh lattice in the increments of $0.01$ from $-\frac{1}{2}$ to $\frac{1}{2}$. A hint vector $\mathbf{h}$ is good or poor, depending on the secret bits $x_\kappa$; for instance, $h_0 = h_1 = \frac{1}{2}$ is the best hint in case of $x_0=x_1=0$, while it is the worst in case of $x_0 = x_1 = 1$. These hold for both of cDM and qDM. Bob's average scores are undifferentiated in both DMs at each corner point, whereas they differentiate, if far from the corners, maximally near to the origin. At the origin, both DMs have score value of $0$. In the cDM, Bob's average score is continuous on the entire hint space. In the qDM, to the contrary, it is discontinuous as crossing the axes, in particular the origin. The blue and red dashed lines represent the hint vectors with equal degrees $\abs{h_0}=\abs{h_1}$, connecting the minimal and the maximal scores.}
\end{figure*}

{\em Experimental demonstration.}---Now, we design linear-optical settings for the proof-of-principle experiments, as drawn in Fig.~\ref{fig:exp}. To simulate the qDM algorithm, we use single-photon light as the ancillary system input\cite{Naruse15}. Horizontal and vertical polarizations of the photon represent the qubit signal, such that $\ket{H} \leftrightarrow \ket{0}$ and $\ket{V} \leftrightarrow \ket{1}$. The unitary operations $u_j$ ($j=0,1$) can be realized as combinations of half-wave-plate (HWP) and quarter-wave-plate (QWP). More specifically, $u_0$ is composed of HWP($\vartheta_0$)-QWP($\varphi_0$)-QWP($\chi$), and $u_1$ is realized by one HWP($\vartheta_1$). Here, $\vartheta_0$, $\varphi_0$, and $\vartheta_1$ are controllable rotation angles of the wave plates. The angle $\chi$ is fixed to be $\frac{\pi}{4}$. Such a setting for qDM can generate all possible outputs for Bob's guesses by controlling the wave plate angles, according to the following rules:
\begin{eqnarray}
\left\{
\begin{array}{l}
\vartheta_0 = \frac{1}{2}\left( \frac{\Delta}{2} + \cos^{-1}\sqrt{P(u_0 \to \openone)}\right), \\
\varphi_0 = \frac{1}{2}\left( \Delta - \frac{\pi}{2} \right), \\
\vartheta_1 = \frac{1}{2}\cos^{-1}\sqrt{P(u_1 \to \openone)}.
\end{array}
\right.
\label{eq:control_rule_qDM}
\end{eqnarray}
We then also simulate the cDM algorithm for comparison. For cDM, we prepare the thermal state of light as the ancilla input, leaving no room for unexpected quantum effects on the cDM. The signal bits are also represented by the light polarization, i.e., $H \leftrightarrow 0$ and $V \leftrightarrow 1$. However, in such a cDM, application of the given hint $\mathbf{h}$ is limited without the ability to fully exploit the quantum superposition; i.e., the directional information of $\mathbf{h}$ cannot be encoded. The classical operations $u_j$ ($j=0,1$) can thus be implemented with only HWPs placed at either $\vartheta_j=0$ (for $u_j \to \openone$) or $\theta_j=\frac{\pi}{4}$ (for $u_j \to X$), probabilistically, based on Eq.~(\ref{eq:preferences}) (see Fig.~\ref{fig:exp}{\bf b}).

\begin{figure}
\centering
\includegraphics[width=0.46\textwidth]{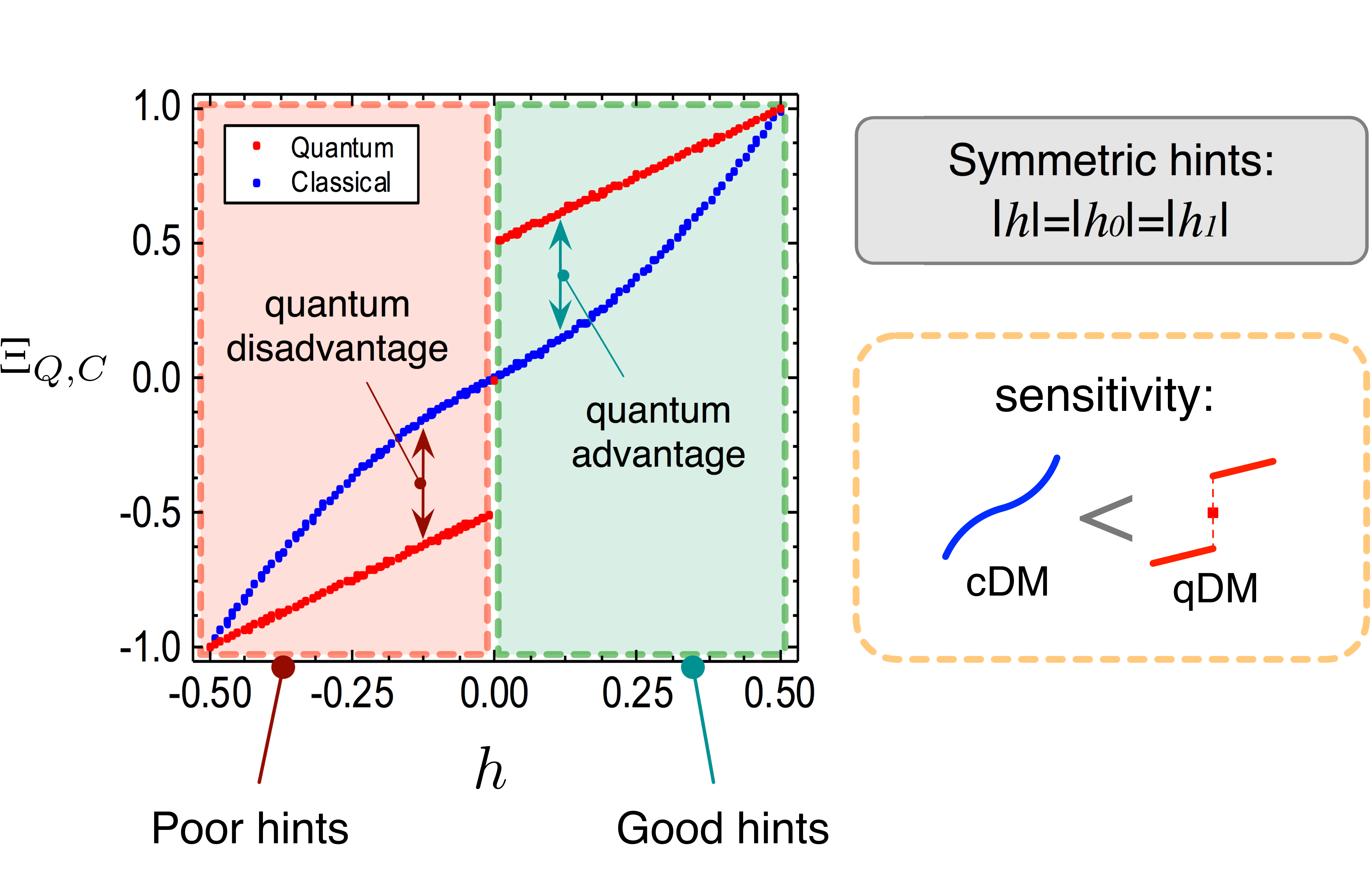} 
\caption{\label{grp:exp_results1-2} {\bf Average scores with respect to the symmetric hints.} The experimentally obtained average scores of Bob are presented along the blue and red dashed lines in Fig.~\ref{grp:exp_results1}{\bf a} and \ref{grp:exp_results1}{\bf b}. These lines, which are toward the best and worst hints from the origin, correspond to the case of symmetric hints, i.e., $\abs{h_0}=\abs{h_1}=\abs{h}$. The red and blue points are Bob's average scores $\Xi_C$ and $\Xi_Q$, respectively, as a function of $h$. Both DMs share the best and the worst scores $\Xi_{Q,C}=\pm 1$ at $h= \pm \frac{1}{2}$, and $\Xi_{Q,C}=0$ at the origin (no hint). For all other points, $\Xi_Q$ is higher (lower) than $\Xi_C$ for good (poor) hints. As a big contrast between cDM and qDM, $\Xi_C$ is continuous in the whole range of symmetric hint $h$, whereas its quantum counterpart $\Xi_Q$ is clearly discontinuous at $h = 0$. $\Xi_Q$ abruptly changes near the origin when the hint h passes the origin, resembling critical phenomena of matters.}
\end{figure}

The experiments are carried out for all of Alice's possible strategies, i.e., her choices of the secret bits $x_0$ and $x_1$. In the experiments, we evaluate Bob's average scores $\Xi_C$ and $\Xi_Q$ by repeating $10^4$ games for a given $\mathbf{h}=(h_0, h_1)^T$. We perform such evaluations by varying $h_0$ and $h_1$ from $-0.5$ to $0.5$ at $0.01$ increments. Thus a given hint $\mathbf{h}$ is good or poor for the secret bits $x_\kappa$, which holds for both in cDM and qDM. We represent the experimental results of $\Xi_C$ and $\Xi_Q$ as density-plots in the space of $h_0$ and $h_1$ (see Fig.~\ref{grp:exp_results1}). The average scores $\Xi_C$ and $\Xi_Q$ are undifferentiated at each corner point, whereas they differentiate, if far from the corners, maximally near to the origin, i.e., when the hints are very small. At the origin, i.e., $h_0=h_1=0$, the average scores are to be zero in both DMs. Here, note that in the qDM, Bob's average score $\Xi_Q$ is discontinuous as crossing the axes, while $\Xi_C$ is continuous everywhere in the cDM. Meanwhile, $\Xi_Q$ is always higher (lower) than $\Xi_C$ for good (poor) hints. To see these features conspicuously, we also perform experiments for the symmetric hints, i.e., $\abs{h_j} = \abs{h}$, along the blue and red dashed lines in Fig.~\ref{grp:exp_results1}{\bf a} and \ref{grp:exp_results1}{\bf b}. These lines, which are toward the best and worst hints from the origin, are represented by $h$ whose sign is positive (negative) when its quality is good (poor). The result clearly shows the abrupt score-change between the quantum advantage $\Gamma$ and disadvantage $-\Gamma$ (see Fig.~\ref{grp:exp_results1-2}). All these results indicate that qDM exhibits higher sensitivity between the boundary for good and poor hints, as described in Eq.~(\ref{eq:qpt}).

\begin{figure}
\centering
\includegraphics[width=0.46\textwidth]{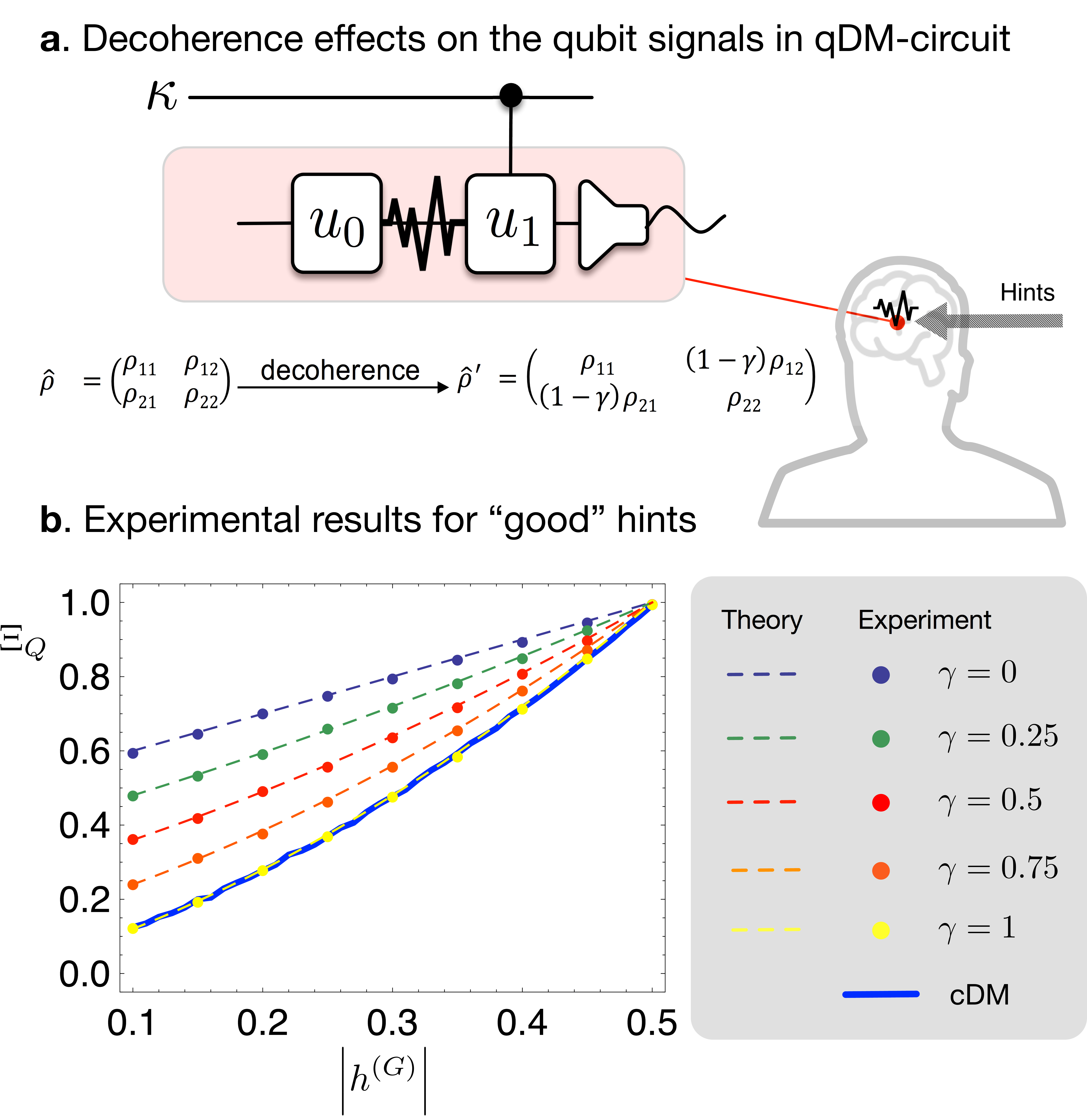} 
\caption{\label{grp:exp_results2} {\bf Decoherence effect on the qDM algorithm.} ({\bf a}) We consider the decoherence effect that arises between the operations $u_0$ and $u_1$ on the ancilla qubit channel for the qDM algorithm. ({\bf b}) Experimental simulations are carried out for different values of the decoherence rate $\gamma$ ($0$ to $1$, $0.25$ step). The symmetric hint $h$ is assumed to be positive, $h>0$. Bob's average scores $\Xi_Q$ are presented for the experimental data (dots) and for the theoretical predictions of qDM (dashed lines) in Eq.~(\ref{eq:decohered_gamma}) together with the experimental data of cDM (blue solid line). The results clearly show that the quantum advantage, i.e., the positive differential from the cDM score decreases as increasing the decoherence rate $\gamma$, and the quantum score eventually becomes equal to the classical if completely decohered with $\gamma=1$.}
\end{figure}

Analyzing further, we consider the decoherence effects, which cause degradation of the quantum superposition, during the process of qDM. Here, without loss of the generality, the signals transmitted in the ancillary system in qDM are assumed to be decohered (mathematically, a decay of off-diagonal elements of the density matrix of the signal state $\hat{\rho}$)\cite{Audretsch08} at a rate of $1-\gamma \le 1$. Then, it is predicted that the decoherence effectively results in a smaller hint-sensitivity with 
\begin{eqnarray}
\Gamma \to \left( 1- \gamma \right)\Gamma. 
\label{eq:decohered_gamma}
\end{eqnarray}
With this prediction, the experiments are carried out for symmetric hints $\abs{h}=\abs{h_0}=\abs{h_1}$. Here, the hints are assumed to be good. The experiments are repeated for $10^4$ games to evaluate the average score $\Xi_Q$. The experimental results clearly confirm the prediction: the quantum advantages become smaller with increasing decoherence rate $\gamma$ (see Fig.~\ref{grp:exp_results2}). However, note that even in this case, qDM still has more advantages than cDM, unless the quantum superposition is completely washed out. This result is also quite remarkable, since quantum properties usually disappear rapidly with very small decoherence.

\section*{DISCUSSION}

We performed the study of quantum decision making, adopting a two-player game where one player (Bob) tries to guess the secret bit numbers chosen by the other player (Alice). In this game, we focused on Bob's decision process in terms of his guesses. Primarily, we attempted to investigate novel quantum features, assuming that Bob (i.e., the decision maker) uses a pre-programmed algorithm by which favorable quantum properties can be exploited. As the main result, we demonstrated both theoretically and experimentally that the quantum aspects make the choosing tendency stronger in the quantum, establishing the high sensitivity at the boundary of opposite hint quality. This quantum feature originates from the fact that quantum DM is able to find additional way of using the quality (i.e., the directional condition) of the given hint $\mathbf{h}$, while the classical DM uses only the amount (i.e., the size). Through the further experiments and analyses, we demonstrated that the high hint-sensitivity persists before the quantum coherence is completely destroyed. Our study is expected to provide the insight to understand some DM processes at the quantum level. 

This work is also intimate to the issue whether novel quantum features exist in a classical game. The issue has been regarded to be negative, while quantum features in quantum games have been discussed mostly by considering quantum strategies\cite{SJvanEnk02,Aharon08}. To attack the issue, on the other hand, we proposed to employ the machinery that plays (or simulates) the decision processes made by the rational players. We hope that the present work would accelerate the studies on potential applications, including quantum cryptography\cite{Werner09,Kaniewski16} and quantum machine learning\cite{Clausen18}.

\section*{METHODS}

{\bf Preparation of the ancillary input.} In the qDM experiments, we prepared a heralded single-photon state ($H$-polarized) as the ancillary input. Photon pairs are produced in type-II spontaneous parametric down conversion (SPDC) using a periodically poled $\text{KTiOPO}_4$ crystal (length, $10$ mm) and a continuous wave pump laser (wavelength, $401.5$ nm). The vertically polarized photons reflected by a PBS are used as trigger photons, and the transmitted horizontally polarized photons are used as signal photons. Signal photons were counted only when the trigger photons were detected. Here, if this post-selection is not applied, the signals toward the gate operations are the thermal state with supper-Poissonian photon statistics. In the cDM experiments, the thermal state of light was employed as the ancillary input, which does not possess the quantum coherence (see Fig.~\ref{fig:exp}).
 
{\bf Experimental simulation of decoherence.} Effectively, the decoherence can be simulated in the experiments by setting the relative phases of the states either as $0$ or as $\pi$ (a phase flip) randomly with a ratio of $1-\gamma/2$ to $\gamma/2$. Then, statistically, the state $\rho$ can be described as\cite{Audretsch08}
\begin{eqnarray}
\hat{\rho} \to \hat{\rho}' = \begin{pmatrix} \rho_{11} & (1-\gamma) \rho_{12} \\ (1-\gamma) \rho_{21} & \rho_{22} \end{pmatrix}.
\end{eqnarray}

\section*{References}

\bibliographystyle{iop}

\begin{thebibliography}{26}

\bibitem{Zsambok14}
Zsambok, C. E., \& Klein, G. (Eds.) Naturalistic decision making. (Psychology Press, 2014).

\bibitem{Tversky74}
Tversky, A., \& Kahneman, D. Judgment under uncertainty: Heuristics and biases. {\em Science} {\bf 185}, {1124--1131} (1974).

\bibitem{Tversky92}
Tversky, A., \& Shafir, E. The disjunction effect in choice under uncertainty. {\em Psychol. Sci.} {\bf 3} {305--310} (1992).

\bibitem{Blackhart05}
Blackhart, G. C., \& Kline, J. P. Individual differences in anterior EEG asymmetry between high and low defensive individuals during a rumination/distraction task. {\em Pers. Individ. Dif.} {\bf 39}, {427--437} (2005). 

\bibitem{Resulaj09}
Resulaj, A., Kiani, R., Wolpert, D. M., \& Shadlen, M. N. Changes of mind in decision-making. {\em Nature} {\bf 461}, {263} (2009).

\bibitem{Julio10} 
Gonz\'{a}lez-D\'{i}az, J., Garc\'{i}a-Jurado, I. \& Fiestras-Janeiro, M. G. An Introductory Course on Mathematical Game Theory, vol. 115 of Graduate Studies in Mathematics (American Mathematical Society, 2010).

\bibitem{Meyer99}
Meyer, D. A. Quantum Strategies. {\em Phys. Rev. Lett.} {\bf 82}, {1052--1055} (1999).

\bibitem{Eisert99}
Eisert, J., Wilkens, M. \& Lewenstein, M. Quantum Games and Quantum Strategies. {\em Phys. Rev. Lett.} {\bf 83}, {3077--3080} (1999).

\bibitem{Lee03}
Lee, C. F. \& Johnson, N. F. Efficiency and formalism of quantum games. {\em Phys. Rev. A} {\bf 67}, {022311} (2003).

\bibitem{Deutsch99}
Deutsch, D. Quantum theory of probability and decisions. {\em Proc. R. Soc. A} {\bf 455}, {3129} (1999).

\bibitem{Pothos09}
Pothos, E. M., \& Busemeyer, J. R. A quantum probability explanation for violations of `rational' decision making. {\em Proc. R. Soc. B} {\bf 276}, {2171} (2009).

\bibitem{SJvanEnk02}
van Enk, S. J. \& Pike, R. Classical rules in quantum games. {\em Phys. Rev. A} {\bf 66}, {024306} (2002).

\bibitem{Aharon08}
Aharon, N. \& Vaidman, L. Quantum advantages in classically defined tasks. {\em Phys. Rev. A} {\bf 77}, {052310} (2008).

\bibitem{Bang16}
Bang, J., Ryu, J., Paw\l{}owski, M., Ham, B. S., \& Lee, J. Quantum-mechanical machinery for rational decision-making in classical guessing game. {\em Sci. Rep.} {\bf 6}, {21424} (2016).

\bibitem{Mihatsch02}
Mihatsch, O., \& Neuneier, R. Risk-sensitive reinforcement learning. {\em Mach. Learn.} {\bf 49}, {267--290} (2002).

\bibitem{Lee08}
Lee, D. Game theory and neural basis of social decision making. {\em Nat. Neurosci.} {\bf 11}, {404} (2008).

\bibitem{Molleman14}
Molleman, L., Van den Berg, P., \& Weissing, F. J. Consistent individual differences in human social learning strategies. {\em Nat. Commun.} {\bf 5}, {3570} (2014).

\bibitem{Ghahramani15}
Ghahramani, Z. Probabilistic machine learning and artificial intelligence. {\em Nature} {\bf 521}, {452} (2015).

\bibitem{Lungo05}
Lungo, A. D., Louchard, G., Marini, C. \& Montagna, F. The Guessing Secrets problem: a probabilistic approach. {\em J. Algorithm.} {\bf 55}, {142--176} (2005).

\bibitem{Lehner13}
Lehner, P. E., Mullin, T. M., \& Cohen, M. S. When Should a Decision Maker Ignore the Advice of a Decision Aid? {\em arXiv preprint arXiv:1304.1515} (2013). 

\bibitem{Park16}
Park, C. Y., Kang, M., Lee, C. W., Bang, J., Lee, S. W., \& Jeong, H. Quantum macroscopicity measure for arbitrary spin systems and its application to quantum phase transitions. {\em Phy. Rev. A} {\bf 94}, {052105} (2016).

\bibitem{Naruse15}
Naruse, M., Berthel, M., Drezet, A., Huant, S., Aono, M., Hori, H., \& Kim, S. J. Single-photon decision maker. {\em Sci. Rep.} {\bf 5}, {13253} (2015).

\bibitem{Audretsch08}
Audretsch, J. Entangled Systems: New Directions in Quantum Physics. (John Wiley \& Sons, 2008).

\bibitem{Werner09}
Werner, A. H., Franz, T., \& Werner, R. F. Quantum cryptography as a retrodiction problem. {\em Phys. Rev. Lett.} {\bf 103}, {220504} (2009).

\bibitem{Kaniewski16}
Kaniewski, J., \& Wehner, S. Device-independent two-party cryptography secure against sequential attacks. {\em New J. Phys.} {\bf 18}, {055004} (2016). 

\bibitem{Clausen18}
Clausen, J., \& Briegel, H. J. Quantum machine learning with glow for episodic tasks and decision games. {\em Phys. Rev. A} {\bf 97}, {022303} (2018).

\end{thebibliography}

\section*{Acknowledgments} 

The authors thank Jaewan Kim and Byoung Seung Ham for valuable discussions. JB thanks Marcin Wie\ifmmode \acute{s}\else \'{s}\fi{}niak, Wies\l{}aw Laskowski, Marcin Paw\l{}owski. This research was supported through the National Research Foundation of Korea (NRF) grant (No. 2014R1A2A1A10050117 and No. 2016R1A2B4014370) and the Institute for Information and communications Technology Promotion (IITP-2018-2015-0-00385), funded by the Korea government(MSIT), Korea.  This research was also implemented as a research project on quantum machine learning (No. 2018-104) by the ETRI affiliated research institute. JB acknowledge the support of the R\&D Convergence program of NST (National Research Council of Science and Technology) of Republic of Korea (No. CAP-18-08-KRISS).


\onecolumngrid
\clearpage

\section{Supplementary Material for ``Quantum Sensitivity to Information Quality in Decision Making''}

\renewcommand{\thesection}{S\arabic{section}}   
\renewcommand{\thetable}{S\arabic{table}}   
\renewcommand{\thefigure}{S\arabic{figure}}
\renewcommand{\theequation}{S\arabic{equation}}

\section{The theoretical analysis of Bob's average payoff (score)}

\subsection{The operations $u_j$ ($j=0,1$) in the classical and quantum decision-making process} 

In our study, Bob's decision-making (DM) process can be described as a function $f: x_\kappa \to m_\kappa$ ($\kappa=0,1$). Here, the measurement outcome $m_\kappa$ is supposed to be the outcomes of Bob's decision, i.e., $m_\kappa \to y_\kappa$, as described in the main manuscript. The function $f$ is defined with the two operations $u_0$ and $u_1$ in the ancillary system. More specifically, it implements four possible functions $f$ depending on the pair ($u_0$, $u_1$), such that 
\begin{eqnarray}
&[{\tau.1}]& (\openone, \openone) \leftrightarrow f_{\tau=1}(x_\kappa) = 0, \nonumber \\
&[{\tau.2}]& (\openone, X) \leftrightarrow f_{\tau=2}(x_\kappa) = x_\kappa, \nonumber \\ 
&[{\tau.3}]& (X, \openone) \leftrightarrow f_{\tau=3}(x_\kappa) = 1, \nonumber \\
&[{\tau.4}]& (X, X) \leftrightarrow f_{\tau=4}(x_\kappa) = 1 \oplus x_\kappa, 
\label{eq:case_tau}
\end{eqnarray}
where $\openone$ and $X$ denote the identity and logical-not operations, respectively. Here, we set $\alpha=0$ for simplicity (see Table in Fig.~2 of the main manuscript). Then, we recall the classical and quantum versions of DM process. As described in our main text, cDM is defined with the classical elements of the ancillary system; the ancilla input $\alpha=0$ is a binary number, and the operations $u_j$ are applied randomly (either to be $\openone$ or to be $X$) based on the preferences $P(u_j \to \openone)$ and $P(u_j \to X)$ ($j=0,1$). Thus, the probabilistic application of $u_j$ is represented by a stochastic evolution matrix,
\begin{eqnarray}
\begin{pmatrix}
P(u_j \to \openone) & P(u_j \to X) \\
P(u_j \to X) & P(u_j \to \openone)
\label{eq:sto_evol_mat}
\end{pmatrix}.
\end{eqnarray}

On the other hand, the qDM is defined with the quantum input $\ket{\alpha}=\ket{0}$ and the application of $u_j$ is represented by a unitary matrix,
\begin{eqnarray}
\begin{pmatrix}
\sqrt{P(u_j \to \openone)} & e^{i \phi_j}\sqrt{P(u_j \to X)} \\
e^{-i \phi_j}\sqrt{P(u_j \to X)} & -\sqrt{P(u_j \to \openone)}
\label{eq:unitary_mat}
\end{pmatrix}.
\end{eqnarray}
which inherently involves (quantum) probabilistic properties. Here, note that the additional degree of freedom, i.e., the quantum phase $\phi_j$ ($j=0,1$), is introduced to faithfully deal with the quantum superposition property. 

\subsection{The calculations of Bob's payoffs} 

One crucial task in game theory is to characterize a function $\$$, which determines the average payoffs of the players over the number of games:
\begin{eqnarray}
\$ : S \times H \to \Xi^{(i)} \in \mathbb{R},
\end{eqnarray}
where $S$ and $H$ denote the set of possible strategies and preferences, respectively. Here, $\Xi^{(i)}$ is the average payoff of the $i$-th player. In our game, Bob's average payoff $\Xi$ can be written, explicitly, as
\begin{eqnarray}
\Xi = \frac{1}{4}\sum_{\tau=1}^4 \overline{\xi}_\tau
\end{eqnarray}
where we assumed that Alice chooses her secret bits $x_\kappa$ at random. The value $\overline{\xi}_\tau$ ($\tau=1,2,3,4$) is defined as the payoff averaged for a specific cases of $\tau$, defined in Eq.~(\ref{eq:case_tau}), i.e., 
\begin{eqnarray}
\overline{\xi}_\tau = \sum_{x_\kappa \in \{0,1\}} \frac{\xi}{2} \Big( P(x_\kappa = y_\kappa) - P(x_\kappa \neq y_\kappa) \Big),
\label{eq:avxi_tau}
\end{eqnarray}
where the index $\tau$ specifies one of the cases [$\tau.1$]-[$\tau.4$]. Here, $P(x_\kappa = y_\kappa)$ and $P(x_\kappa \neq y_\kappa)$ are the probabilities that the outcome of Bob's decision is correct and incorrect for the given $x_\kappa$, respectively. For later analysis, we rewrite Eq.~(\ref{eq:avxi_tau}), for each $\tau$, as below
\begin{eqnarray}
\overline{\xi}_{\tau=1} &=& \frac{\xi}{2} \Big( P(y_0 = 0) + P(y_1 = 0) - P(y_0 = 1) - P(y_1 = 1) \Big), \nonumber \\
\overline{\xi}_{\tau=2} &=& \frac{\xi}{2} \Big( P(y_0 = 0) + P(y_1 = 1) - P(y_0 = 1) - P(y_1 = 0) \Big), \nonumber \\
\overline{\xi}_{\tau=3} &=& \frac{\xi}{2} \Big( P(y_0 = 1) + P(y_1 = 1) - P(y_0 = 0) - P(y_1 = 0) \Big), \nonumber \\
\overline{\xi}_{\tau=4} &=& \frac{\xi}{2} \Big( P(y_0 = 1) + P(y_1 = 0) - P(y_0 = 0) - P(y_1 = 1) \Big),
\label{eq:avxi_sp}
\end{eqnarray}
where $P(y_\kappa = m_\kappa)$ is the probability of choosing the final strategy $y_\kappa = m_\kappa$ ($m_\kappa=0,1$) in our DM algorithm, described in Fig.~2 in the main manuscript.
\begin{figure}[t]
\centering
\includegraphics[width=0.4\textwidth]{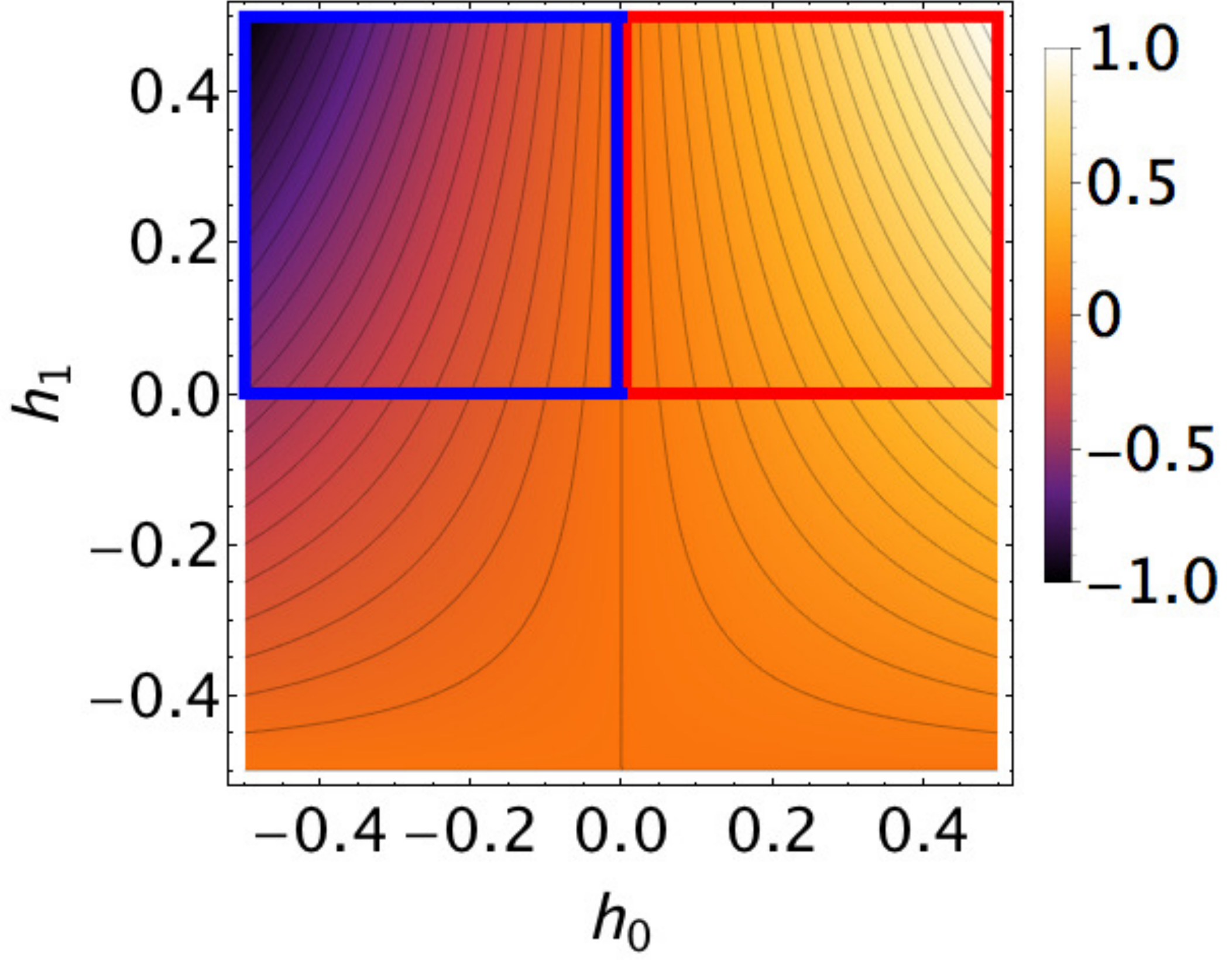}
\includegraphics[width=0.4\textwidth]{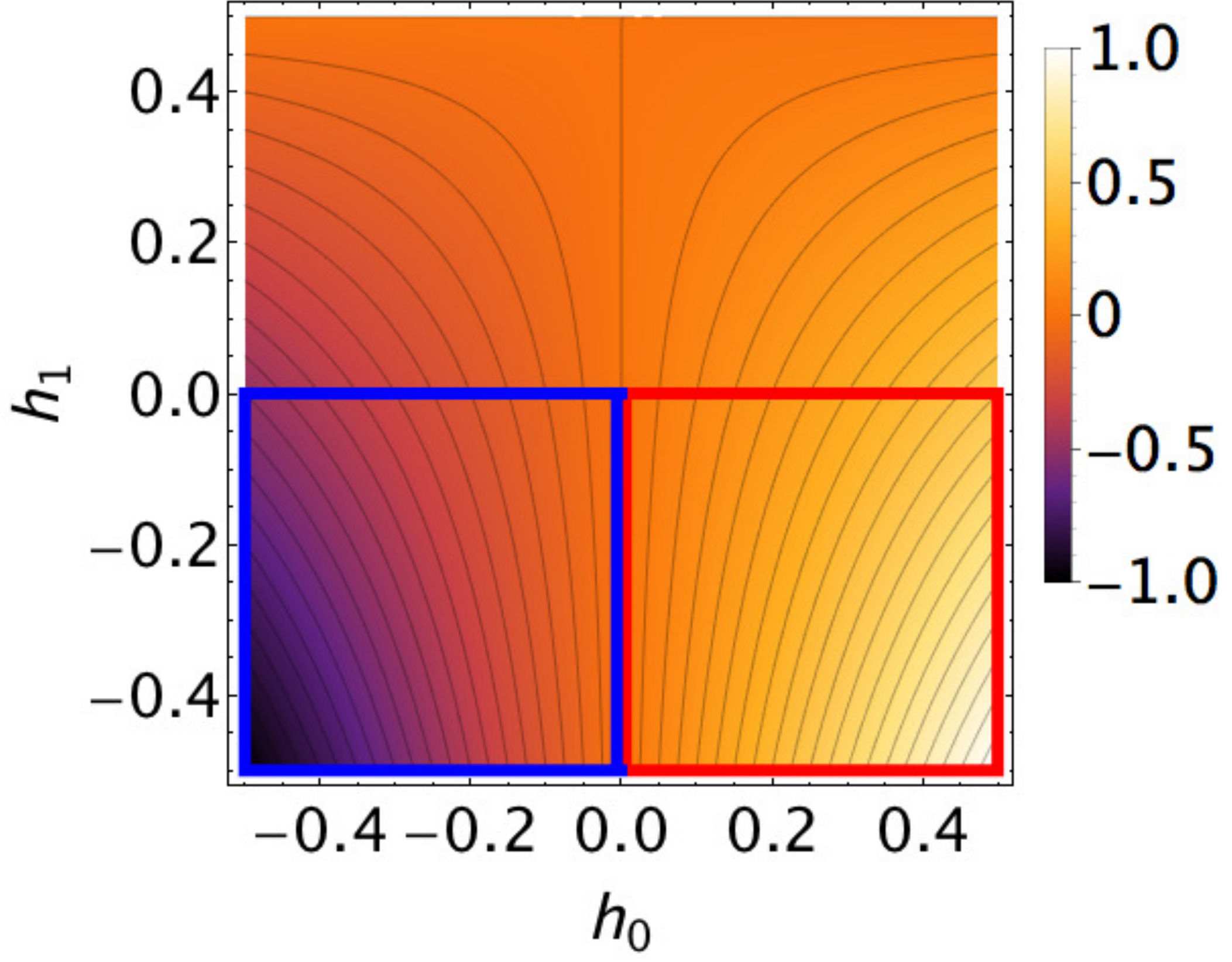}
\\
\includegraphics[width=0.4\textwidth]{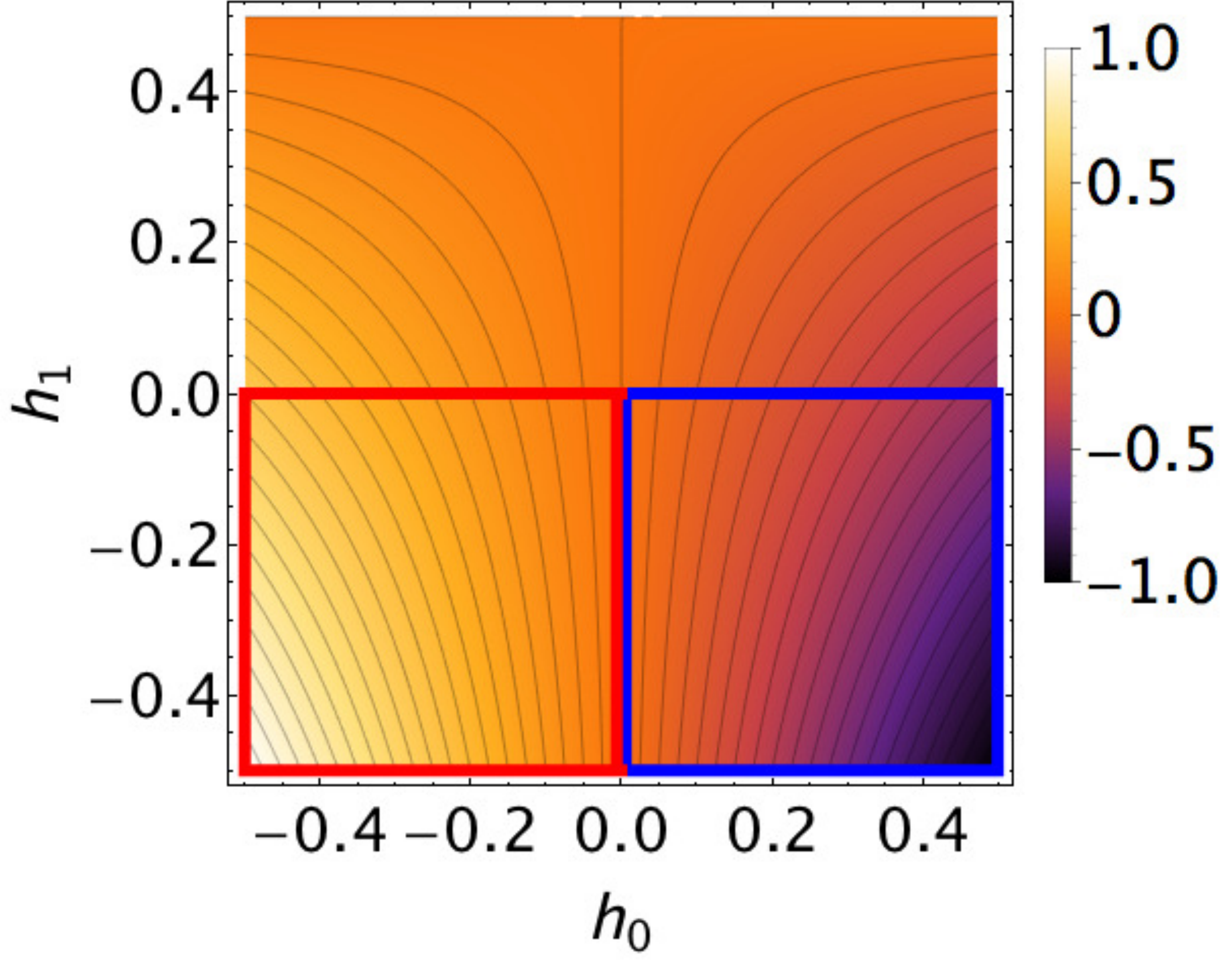}
\includegraphics[width=0.4\textwidth]{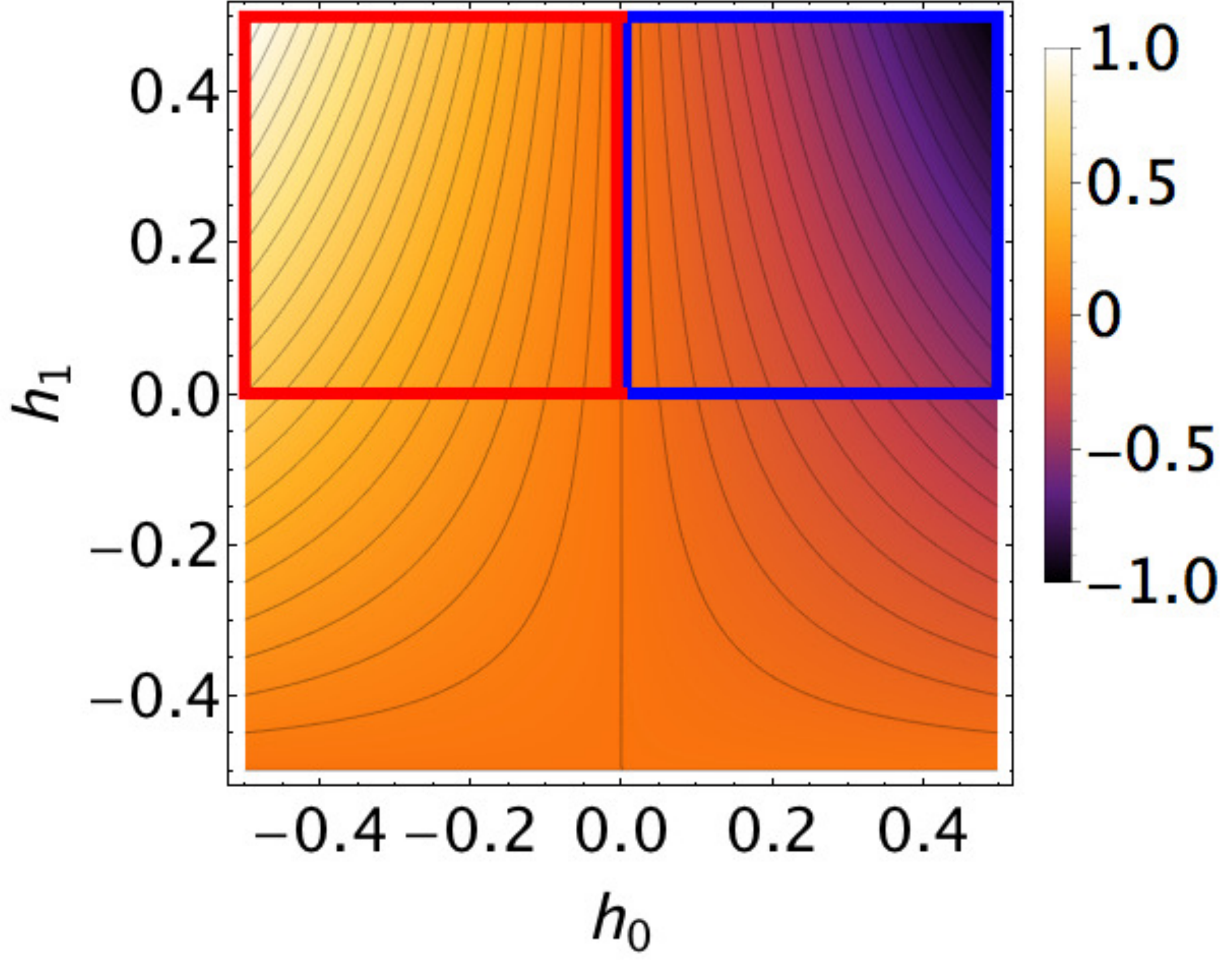}
\caption{{\bf Bob's average payoffs in cDM.} We depict the theoretically expected $\overline{\xi}_{\tau,C}$ averaged for a specific set of $x_\tau \in \{0,1\}$: (top-left) [$\tau.1$], (top-right) [$\tau.2$], (bottom-left) [$\tau.3$], and (bottom-right) [$\tau.4$]. We specify the regions of the good hints (red-line box) and the poor hints (blue-line box) (see, also, Fig.~4 in our main text).}
\label{grp:app_score_Cl}
\end{figure}

{\em 1) Analysis of cDM.} -- Now, we calculate Bob's average payoff $\Xi_C$ achievable from cDM. To do this, let us first write the classical probabilities $P_C(y_\kappa = m_\kappa)$ in Eq.~(\ref{eq:avxi_sp}) in terms of the DM preferences casted in Eq.~(\ref{eq:sto_evol_mat}), such that
\begin{eqnarray}
P_C(y_0 = 0) &=& P(u_0 \to \openone) = \frac{1}{2} + h_0, \nonumber \\
P_C(y_0 = 1) &=& P(u_0 \to X) = \frac{1}{2} - h_0, \nonumber \\
P_C(y_1 = 0) &=& P(u_0 \to \openone)P(u_1 \to \openone) + P(u_0 \to X)P(u_1 \to X) = \frac{1}{2} + 2 h_0 h_1, \nonumber \\
P_C(y_1 = 1) &=& P(u_0 \to \openone)P(u_1 \to X) + P(u_0 \to X)P(u_1 \to \openone) = \frac{1}{2} - 2 h_0 h_1.
\label{eq:probs_out_C}
\end{eqnarray}
Then, we can write $\overline{\xi}_{\tau,C}$ for cDM, using Eqs.~(\ref{eq:avxi_tau})-(\ref{eq:probs_out_C}), as below.
\begin{eqnarray}
\overline{\xi}_{\tau=1,C} &=& h_0 + 2 h_0 h_1, \nonumber \\
\overline{\xi}_{\tau=2,C} &=& h_0 - 2 h_0 h_1, \nonumber \\
\overline{\xi}_{\tau=3,C} &=& -h_0 + 2 h_0 h_1, \nonumber \\
\overline{\xi}_{\tau=4,C} &=& -h_0 - 2 h_0 h_1,
\label{eq:pc_x}
\end{eqnarray}
where the constant $\xi$ is assumed to be $1$ without loss of the generality. Here, it is obvious that if there is no bias among the preferences, i.e., no hints are provided as $h_0=h_1=0$, then $\overline{\xi}_{\tau,C}=0$ for all $\tau=1,2,3,4$. However, if Bob has non-zero hints $\mathbf{h}=(h_0, h_1)^T$, Bob can improve his winning average with good hint $\mathbf{h}$. Here, by ``good'' we mean that the directional conditions of $\mathbf{h}$ is appropriately assigned toward $(x_0, x_1)^T$. More specifically, Bob can have
\begin{eqnarray}
\Xi_C = \abs{h_0} + 2 \abs{h_0} \abs{h_1}.
\label{eq:xiC_best}
\end{eqnarray}
However, if the hint is poor, Bob may fail. In particular, we can imagine the worst case that any malicious hint misleads Bob, in which Bob will have the payoff
\begin{eqnarray}
\Xi_C = -\abs{h_0} - 2 \abs{h_0} \abs{h_1}. 
\label{eq:xiC_worst}
\end{eqnarray}
To see this clearly, we draw the graphs of $\overline{\xi}_{\tau,C}$ for $\tau=1,2,3,4$ based on the theoretical analysis (see Fig.~\ref{grp:app_score_Cl}). In each graph, we specify the regions of the good hints (red-line box) and the poor hints (blue-line box) in the space of ($h_0$, $h_1$). This is well matched to our experimental results in Fig.~4 of the main manuscript.

\begin{figure}[t]
\centering
\includegraphics[width=0.4\textwidth]{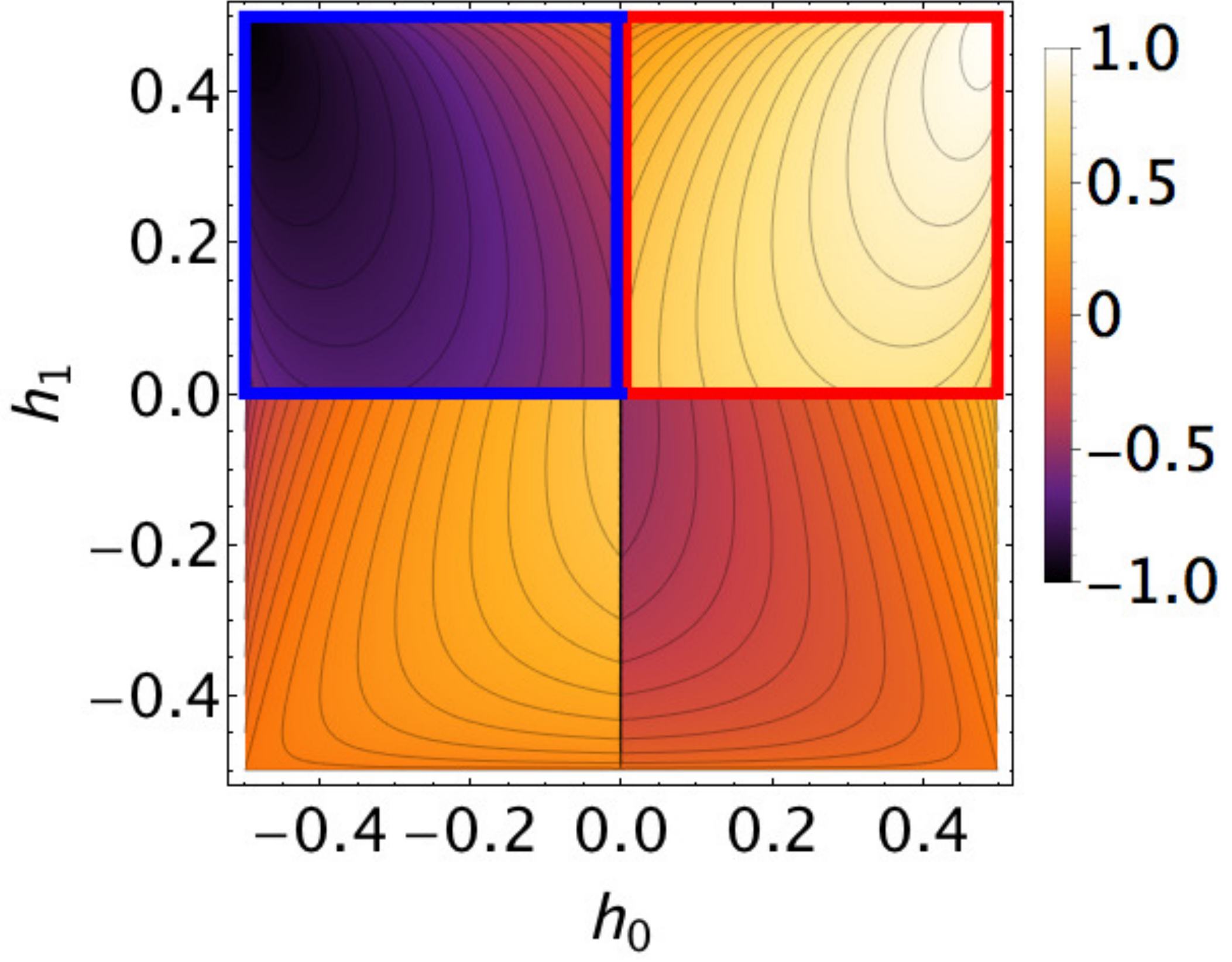}
\includegraphics[width=0.4\textwidth]{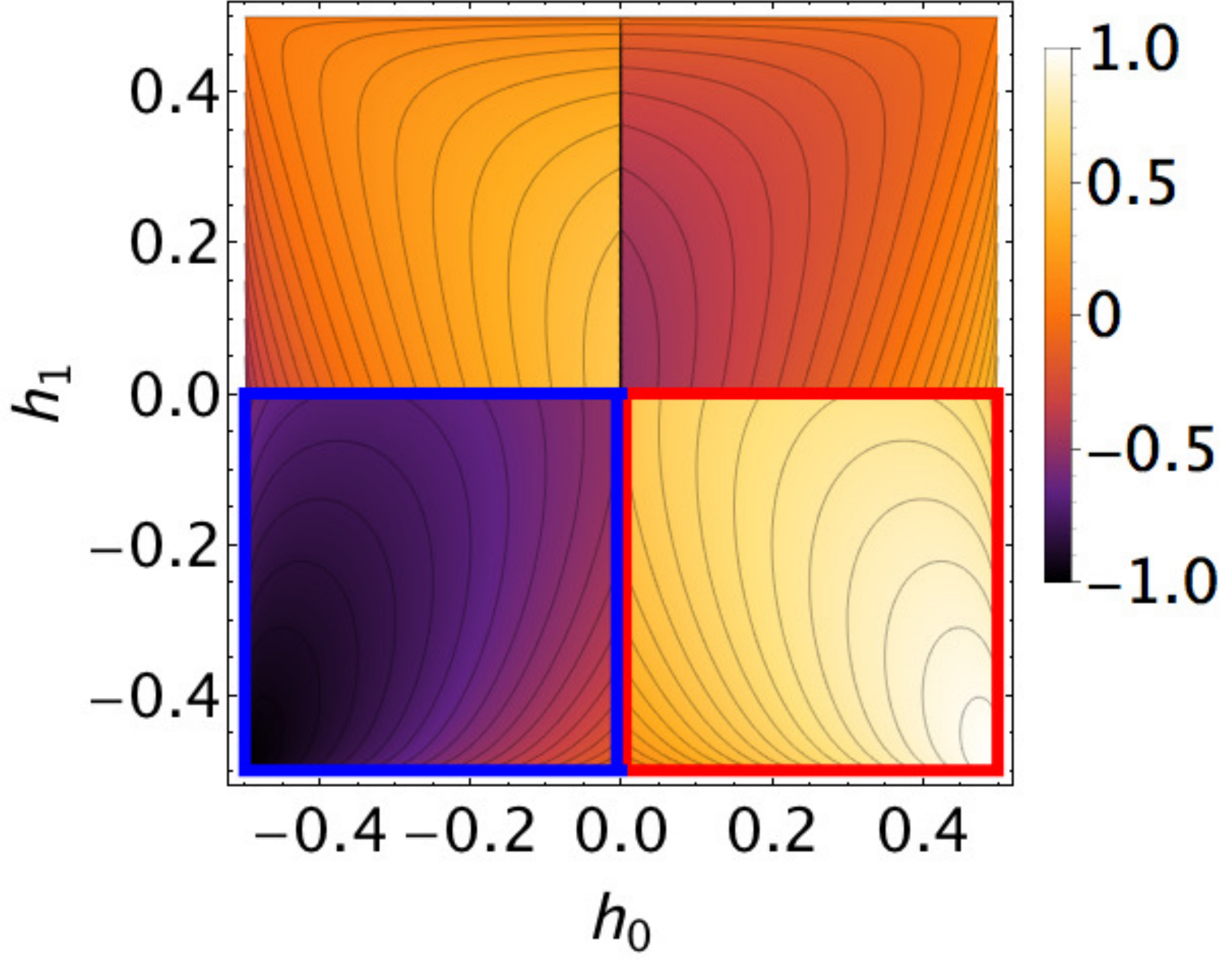}
\\
\includegraphics[width=0.4\textwidth]{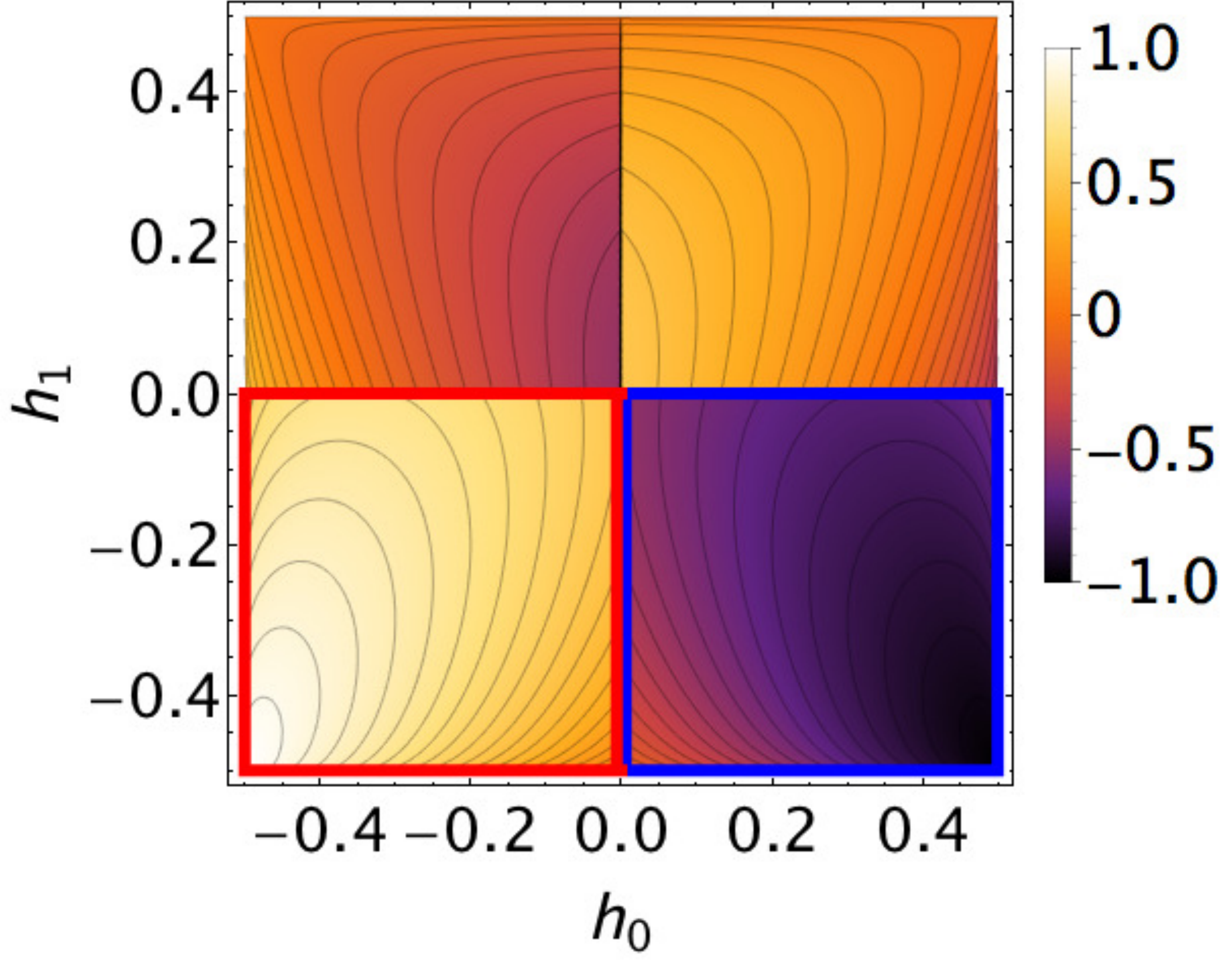}
\includegraphics[width=0.4\textwidth]{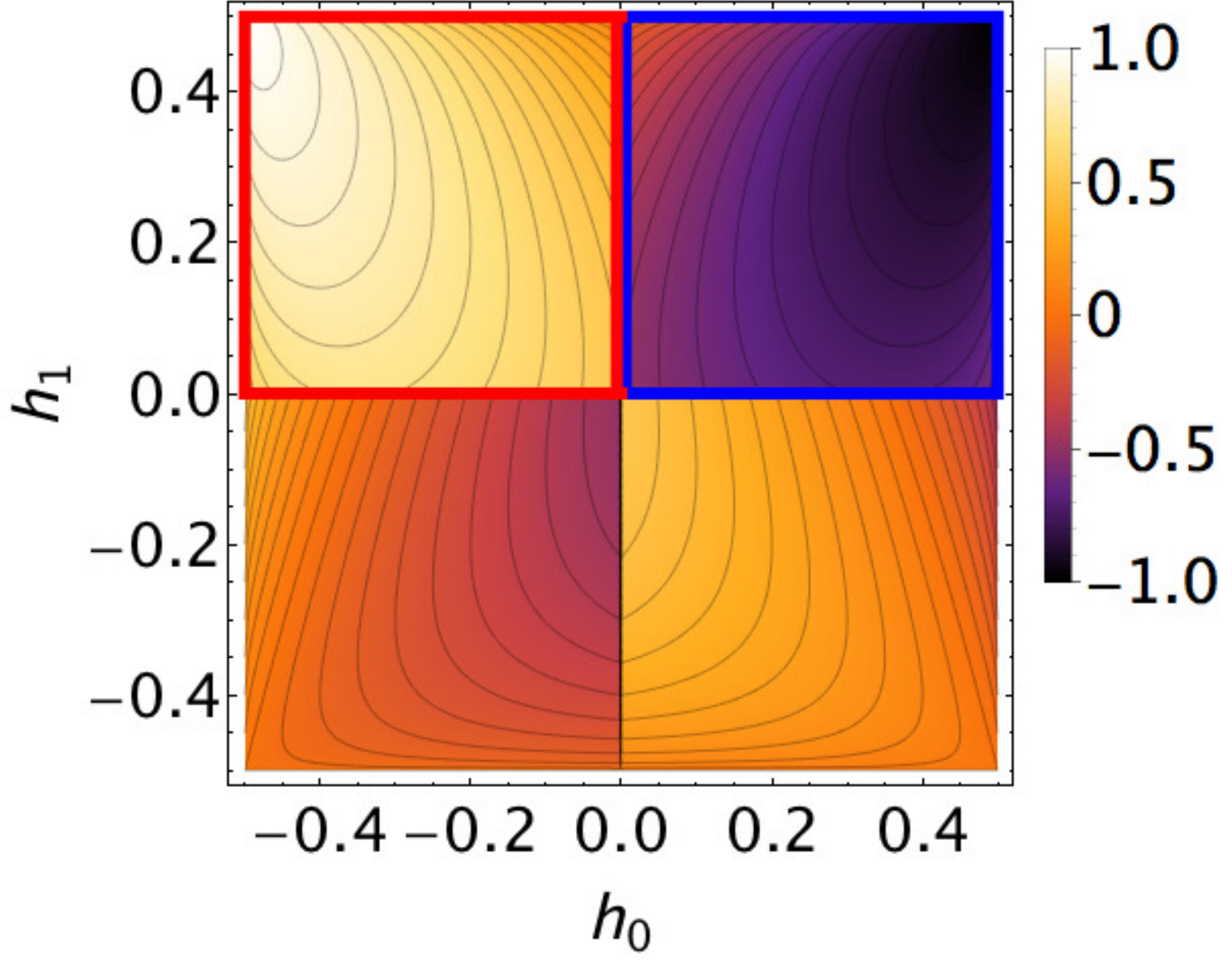}
\caption{{\bf Bob's average payoffs in qDM.} We depict the theoretically expected $\overline{\xi}_{\tau,Q}$ for (top-left) [$\tau.1$], (top-right) [$\tau.2$], (bottom-left) [$\tau.3$], and (bottom-right) [$\tau.4$]. We also specify the regions of the good hints (red-line box) and the poor hints (blue-line box) (see, also, Fig.~4 in our main text).}
\label{grp:app_score_Qu}
\end{figure}

{\em 2) Analysis of qDM.} -- Turning our analysis to the qDM, let us first write the quantum version of the probabilities $P_Q(y_\kappa = m_\kappa)$ as below
\begin{eqnarray}
P_Q(y_0 = 0) &=& \abs{\bra{m_0 = 0}\hat{u}_0\ket{\alpha}}^2 = P(u_0 \to \openone) = \frac{1}{2} + h_0, \nonumber \\
P_Q(y_0 = 1) &=& \abs{\bra{m_0 = 1}\hat{u}_0\ket{\alpha}}^2 = P(u_0 \to X) = \frac{1}{2} - h_0, \nonumber \\
P_Q(y_1 = 0) &=& \abs{\bra{m_0 = 0}\hat{u}_1 \hat{u}_0 \ket{\alpha}}^2 = P(u_0 \to \openone)P(u_1 \to \openone) + P(u_0 \to X)P(u_1 \to X) \nonumber \\
    &=& \frac{1}{2} + 2 h_0 h_1  + \Gamma\cos{(\pi\Delta)}, \nonumber \\
P_Q(y_1 = 1) &=& \abs{\bra{m_0 = 1}\hat{u}_1 \hat{u}_0 \ket{\alpha}}^2 = P(u_0 \to \openone)P(u_1 \to X) + P(u_0 \to X)P(u_1 \to \openone) \nonumber \\
    &=& \frac{1}{2} - 2 h_0 h_1 - \Gamma\cos{(\pi\Delta)},
\label{eq:probs_out_Q}
\end{eqnarray}
where $\Delta = \abs{\phi_1 - \phi_0}$ and $\hat{u}_{0,1}$ denotes the unitary operation of $u_{0,1}$ in Eq.~(\ref{eq:unitary_mat}). Here, $\Gamma$ is given as [see Eq.~(6) of the main text]
\begin{eqnarray}
\Gamma = 2\sqrt{\left(\frac{1}{4}-\abs{h_0}^2\right)\left(\frac{1}{4}-\abs{h_1}^2\right)}. 
\label{eq:Gamma}
\end{eqnarray}
Then, we can write $\xi_{\tau,Q}$, using Eq.~(\ref{eq:avxi_tau}), Eq.~(\ref{eq:avxi_sp}) and Eq.~(\ref{eq:probs_out_Q}), as
\begin{eqnarray}
\overline{\xi}_{\tau=1,Q} &=& \overline{\xi}_{\tau=1,C} + \Gamma\cos{(\pi\Delta)}, \nonumber \\
\overline{\xi}_{\tau=2,Q} &=& \overline{\xi}_{\tau=2,C} - \Gamma\cos{(\pi\Delta)}, \nonumber \\ 
\overline{\xi}_{\tau=3,Q} &=& \overline{\xi}_{\tau=3,C} + \Gamma\cos{(\pi\Delta)}, \nonumber \\
\overline{\xi}_{\tau=4,Q} &=& \overline{\xi}_{\tau=4,C} - \Gamma\cos{(\pi\Delta)}, 
\label{eq:pq_x}
\end{eqnarray}
where $\xi$ is also assumed to be $1$ and $\Delta$ is determined by Eq.~(4) in the main manuscript. Here, it is also true that Bob cannot improve his winning chance when $h_0 = h_1 =0$; i.e, Bob has $\overline{\xi}_{\tau,Q}=0$ for all $\tau=1,2,3,4$. However, it can be found from Eq.~(\ref{eq:pq_x}) that the average payoff in the qDM can be higher than those in the cDM by $\Gamma$ when provided by a proper value of $\Delta$ (good hint); 
\begin{eqnarray}
\Xi_Q = \Xi_C + \Gamma,
\label{eq:xiQ_best}
\end{eqnarray}
as described also in the main text. However, there can also be malicious hinting, in which case Bob may fail, similarly to the classical case. From the same analysis as in the case of the cDM, we can see that Bob's average Payoff can be decreased. Notably, in the worst case, such disadvantages can be maximized as
\begin{eqnarray}
\Xi_Q = \Xi_C - \Gamma,
\label{eq:xiQ_worst}
\end{eqnarray}
This implies that the qDM can make the situation worse. To see this, let us see the theoretical graphs of $\overline{\xi}_{\tau,Q}$ in Fig.~\ref{grp:app_score_Qu}, where the regions of the good hints (red-line box) and the poor hints (blue-line box) are also specified.

\end{document}